# Integrating quantum-dots and dielectric Mie resonators: a hierarchical metamaterial inheriting the best of both

*Antonio Capretti\*, Arnon Lesage and Tom Gregorkiewicz*

University of Amsterdam, Science Park 904, 1098XH Amsterdam, Netherlands

**ABSTRACT.** Nanoscale dielectric resonators and quantum-confined semiconductors have enabled unprecedented control over light absorption and excited charges, respectively. In this work, we embed luminescent silicon nanocrystals (Si-NCs) into a 2D array of $SiO_2$ nanocylinders, and experimentally prove a powerful concept: the resulting metamaterial preserves the radiative properties of the Si-NCs and inherits the spectrally-selective absorption properties of the nanocylinders. This hierarchical approach provides increased photoluminescence (PL) intensity obtained without utilizing any lossy plasmonic components. We perform rigorous calculations and predict that a freestanding metamaterial enables tunable absorption peaks up to 50% in the visible spectrum, in correspondence of the nanocylinder Mie resonances and of the grating condition in the array. We experimentally detect extinction spectral peaks in the metamaterial, which drive enhanced absorption in the Si-NCs. Consequently, the metamaterial features increased PL intensity, obtained without affecting the PL lifetime, angular pattern and extraction efficiency. Remarkably, our best-performing metamaterial shows +30% PL intensity achieved with a lower amount of Si-NCs, compared to an equivalent planar film without nanocylinders, resulting in a 3-fold average PL enhancement per Si-NC. The principle demonstrated here is general and the Si-NCs can be replaced with other semiconductor quantum dots, rare-earth ions or organic molecules. Similarly, the dielectric medium can be adjusted on purpose. This spectral selectivity of absorption paves the way for an effective light down-conversion scheme to increase the efficiency of solar cells. We envision the use of this hierarchical design for other efficient photovoltaic, photo-catalytic and artificial photosynthetic devices with spectrally-selective absorption and enhanced efficiency.

Nanoscale solid-state systems are boosting the prospects of material science by providing unparalleled optical and electronic functionalities. Semiconductors with size comparable or smaller than the exciton Bohr radius (a few *nm*), such as quantum dots and nanowires, display quantum-confinement properties. The confinement of the electron wavefunction determines the quantization of the energy states, accompanied by the widening of the semiconductor bandgap, the increase of radiative recombination and the onset of carrier multiplication.[1–3] A number of pioneering works have investigated the interplay between plasmonic resonators and quantum dots, showing that this interaction features enhanced absorption and emission rate.[4–6] Unfortunately, plasmonic resonances come with large parasitic losses due to dissipation in the metal.[7,8] This drawback is particularly detrimental for energy-related and -sustainable applications, which require technological solutions excluding lossy metals and expensive/toxic semiconductors.

On the other hand, dielectric nanoparticles with size of hundreds *nm* support Mie resonances without utilizing metal components.[9,10] Differently from plasmonic nanoparticles, whose scattering is dominated by electric modes, their resonances are both magnetic and electric in nature.[8,11] Moreover, they guarantee reduced optical losses with respect of their plasmonic counterpart.[7,8] Interference effects between magnetic and electric dipolar modes have been shown to enable directional light scattering, as well as scattering suppression.[12,13] Building on the spectrally-selective optical properties of the dielectric scatterers,[11,14] dielectric metamaterials have been very recently investigated for controlling the transmission/reflection ratio in the infrared,[15] for light concentration to an absorbing substrate,[16] for nonlinear optical applications,[17] and for flat optics,[18,19] and are currently being explored for photovoltaics.[20,21]

Here, we explore the opportunity to combine quantum dots and dielectric resonators as building blocks of a hierarchical metamaterial, which not only inherits the intrinsic optical and electronic properties of its nanoscale constituents, but also features enhanced performance. This simple yet powerful scheme is of great impact for applications in photovoltaics, photocatalysis and artificial photosynthesis, as well as for increasing the sustainability of the available optoelectronic devices. In this work, we integrate quantum-confined luminescent silicon nanocrystals (Si-NCs) within $SiO_2$ nanocylinders, arranged in a 2D array as depicted in Fig. 1(a-c). We demonstrate that the resulting metamaterial preserves the Si-NC radiative recombination properties, inherits the nanocylinder spectrally-selective absorption, and features enhanced emission intensity. As a proof of concept, we tackle the spectral requirements of a light down-converter for solar cells,[22] namely: i) total absorption of photons with energy $E > 2 E_{gap}$ and ii) total transmission of photons with $E < 2 E_{gap}$, where $E_{gap}$ is the cell bandgap ($E_{gap} \sim 1.11\ eV$ for Si). We address these requirements by designing a hierarchical metamaterial optimized on three levels:

a) Si-NCs are quantum dots made of crystalline Si (see HRTEM in Fig. 1(d)) with typical diameter < 10 nm. The electron wavefunction is broadened in the momentum space due to the Heisenberg uncertainty principle, increasing the rate of radiative recombination. Therefore, the constraint of momentum conservation is relaxed (Si is an indirect semiconductor) and Si-NCs show excitonic photoluminescence (PL) in the red/infrared spectrum.[3,23] Notably, the occurrence of carrier multiplication in ensembles of Si-NCs, due to a process named space-separated quantum cutting (SSQC),[24,25] enables the down-conversion of 1 high-energy photon into 2 photons with half of the original energy. This has the huge potential to enable sunlight spectral shaping and to truly improve the efficiency of solar cells by down-conversion.[22,26–28]

b) The dielectric nanocylinders, shown in the SEM of Fig. 1(e), are made of a solid-state dispersion of Si-NCs in a $SiO_2$ matrix (Si-NC:$SiO_2$ , in brief) and defined by electron-beam lithography. The size of Si-NCs (~2.5 nm) is 2 orders of magnitude smaller than the wavelength of visible light and, therefore, Si-NC:$SiO_2$ is effectively a homogeneous medium. The effective refractive index $n_{eff}$ is a function of the volume fraction of Si-NCs, and it is $n_{eff}\sim1.9$ for the samples produced in this work (see Methods section). The excellent scattering properties of the nanocylinders depend on their size, shape and refractive index $n_{eff}$, enabling the onset of Mie resonances in the visible spectrum.[16,20,21]

c) The 2D hexagonal array of scattering nanocylinders forms an optically-thin metamaterial, as shown in Fig. 1(f). Superabsorption in a thin 1D dielectric metamaterial has been recently shown to arise from the interference between the scattered and the incident waves, which enables an optimal match of the optical impedances.[29] This results into a maximum absorption of 50% for freestanding dielectric metamaterials like their metal counterpart, and 100% absorption with the support of a back-reflector.

In the following, we experimentally demonstrate that the investigated metamaterial supports extinction peaks in correspondence of the nanocylinder Mie resonances and of the grating condition of the array, whose spectral positions are tailored by geometrical design parameters. Our numerical predictions show that the metamaterial absorption can reach up to 50%. The optical losses in $SiO_2$ are negligible in the visible spectrum and, therefore, the main contribution to the absorption comes from optical transitions in the Si-NCs. We experimentally prove that that the increased absorption directly couples to the Si-NCs, resulting in a 3-fold average PL enhancement per Si-NC. Specifically, our experiments indicate that the PL intensity from the best performing metamaterial is +30% compared to a planar Si-NC:$SiO_2$ film (without the nanocylinder patterning), despite the number of photoluminescent Si-NCs is reduced to 43%.

The investigated metamaterial can find application as an optically-thin down-converter of light on the front surface of commercial solar cells. It is fair to remark that this application necessarily requires PL quantum yield (QY) above 50%, while the maximum value achieved so far for Si-NC:$SiO_2$ is 35%.[30] A florid and growing field of research is tackling the increase of QY to match the requirements for applications.[30,31] Nevertheless, the fundamental goal of this work is to demonstrate that the integration of quantum emitters into dielectric resonators enables spectrally-selective enhancement of photon absorption, obtained without affecting the emitter radiative properties, and resulting in an overall increase of PL intensity. Although we propose and use a specific material platform, the demonstrated principle is general and can be applied to other semiconductor quantum-dots and emitting species, such as rare-earth ions and organic molecules. We envision the application of this principle to applications such as light-conversion and spectral shaping for photovoltaics, but also photocatalysis and artificial photosynthesis.

**Methods.** In the deposited Si-NC:$SiO_2$ films, the size of Si-NCs is 2 orders of magnitude smaller than the wavelength of visible light and their volume fraction $f$ is 17±4%. Therefore, Si-NC:$SiO_2$ can be effectively approximated as a homogeneous medium, with intermediate optical properties between Si and $SiO_2$. We apply the Maxwell-Garnett homogenization formula to estimate the effective refractive index $n_{eff}$ in terms of the dielectric matrix, the Si inclusions and volume fraction.[32] The resulting $n_{eff}$ is shown in Supporting Figure 1, and it is in good agreement with the measured transmission and reflection data. The volume fraction $f$ affects both the optical and electronic properties of the Si-NC:$SiO_2$ medium. Any change of $f$ modifies the effective refractive index of the Si-NC:$SiO_2$ medium, according to the homogenization models: as the concentration of Si increases, the refractive index becomes higher. Moreover, as $f$ increases, the average distance between the Si-NCs is reduced and the probability of SSQC increases as well. Our group demonstrated the existence of an optimal distance as far as the probability of SSQC is concerned.[33] Furthermore, the Si-NC size has a strong influence on the QY and an optimum size exists.[30] Therefore, the interplay between $f$, the deposition parameters and the film thickness determines the resulting PL spectrum, QY and SSQC efficiency.[34,35]

We fabricate three sets of metamaterials with nominal nanocylinder height H=100nm, 300nm and 450nm. For each height, we produce (on the same substrate) 6 arrays with nanocylinder diameter from D=164nm to 420nm. A complete list of investigated samples is provided in the Supporting Table 1. In order to have a fair comparison between metamaterials with the same height, we keep the number of Si-NCs equal in all the metamaterial geometries by fixing the nanocylinder area coverage (AC~43%). This choice also determines the array spacing. Moreover, all the samples with the same height, including the planar film used as reference, are made starting from the same deposited Si-NC:$SiO_2$ layer, and therefore the optical and electronic properties are the same. The fabrication process consists of a sequence of eight steps (a-h), as illustrated in Supporting Figure 2. First, we deposit a planar layer made of Si-NC:$SiO_2$ with thickness H=100nm, 300nm and 450nm (on different substrates, steps (a-c)). Then, a nano-lithographic method is used to pattern the metamaterial geometry into the Si-NC:SiO$_2$ layers (steps (d-h)). The fabrication process results in hexagonal 2D arrays of Si-NC:SiO$_2$ nanocylinders. The arrays have a size of 30μm. The resulting pillar height $H$ is equal to the thickness of the original layer deposited in steps (a-c).

The optical extinction is measured in a Zeiss Axio Observer inverted microscope. The light source is a halogen lamp focused by a condenser with numerical aperture NA=0.35, and the transmitted light is collected through a NA=0.75 100x objective. The spectra are recorded by a Princeton Instruments Acton SpectraPro SP2300 spectrometer equipped with a PyLoN:400 (1340 x 400) cryogenically-cooled charge-coupled device (CCD). For the $0^{th}$ order extinction, we use a customized system where the light source is a fiber-coupled lamp focused by a NA=0.42 lens. The transmitted and reflected light beams are collected by identical lenses (for the reflection we additionally use a half transparent mirror) and delivered to an OceanOptics USB2000+Vis-NIR Spectrometer. The PL experiments are performed in the same Zeiss Axio Observer inverted microscope. The excitation is the 488nm line of a Spectra-Physics Stabilite 2017 Ar laser, focused onto a fixed spot of 50μm by a 100x objective with NA=0.75. The excitation power is fixed at 2.5mW. The PL signal is collected by the same objective by using a dichroic mirror, and measured by the spectrometer and the CCD camera. The values reported in this manuscript are the PL intensities integrated across the emission spectrum. For the PL excitation experiment, we use the 415-537.5nm signal of a SOLAR LP603-I Optical Parametric Oscillator (OPO), pumped by the third harmonic of a Nd:YAG LQ629-10 1064nm laser with 100Hz repetition rate and 12ps pulse duration. The OPO signal is collimated and coupled to the inverted microscope, exciting the samples perpendicularly to their surface. For the PL lifetime measurements, we use the CW output at 445nm of a Becker & Hickl diode laser, modulated by a pulse train with 1 kHz repetition frequency and 100 µs pulse duration. The detector is an ID100 Quantique silicon avalanche photodiode with 40 ps timing resolution, connected to a DPC-230 Becker & Hickl timing card. Absolute QY measurements are performed for the planar Si-NC:SiO$_2$ films before nanopatterning. The samples are placed in an integrating sphere (7.5cm in diameter, Newport) with the PL directed into a Solar LS M266 spectrometer coupled with a Hamamatsu S7031- 1108S Vis-CCD camera. For excitation, we used a 150W Hamamatsu L2273 xenon lamp coupled to double grating Solar MSA130 monochromator. The measured QY is ~2%, while the maximum value achieved so far in literature is 35%,[30] that requires a time-consuming optimization of the material deposition process (the absolute value of QY is not relevant for the concept demonstrated in this manuscript). The homebuilt Fourier microscopy setup consists of a 100x objective with NA=0.90, a set of telescope lenses L1 and L2 with equal focal lengths (f = 20 mm) and a Fourier lens L3 (f = 200 mm). The detector is an Andor Technology silicon CCD. The excitation is a PicoQuant LDH diode laser at 450nm and a pinhole is used to selectively excite one metamaterial sample per each acquisition.

We use the transition matrix method, also known as null-field, to calculate the scattering cross-section spectra of individual nanocylinders.[36] In this method, the expansion coefficients for the scattered field are retrieved by combining the null-field equation with the boundary conditions. We also use the rigorous coupled wave approach, also known as Fourier modal method, to calculate the transmission (T), reflection (R) and absorption (1-T-R) spectra of the 2D arrays of nanocylinders.[37] This method is particularly suitable for simulating light interaction with periodic layered structures which are invariant in the direction normal to the periodicity, due to its Fourier basis representation.

**Results and discussion.** The metamaterial transmission, scattering and absorption depend on the nanocylinder size and distance. We quantify this dependence by measuring the extinction spectra in the 350 - 600 nm spectral range, as showed in Fig. 2(a-c) for the nanocylinder height H=100nm, 300nm and 450nm. The extinction is defined as 1-T, with T being the transmittance collected with an angular aperture ±49°, which comprises also significant contributions from diffuse scattering. The extinction spectra of the three homogeneous Si-NC:SiO$_2$ films are conveniently reported in black, and are used as reference samples throughout the manuscript. For very thin metamaterials (H=100nm, Fig. 2(a)), an extinction band appears with increasing diameter D, and the amplitude increases from 5% to only 20% at λ≈400nm. For metamaterials with H=300nm (Fig. 2(b)), the maximum extinction amplitude is boosted up to 70%. In particular, for diameters from 215 nm to 372nm, we can distinguish the occurrence of broad extinction peaks in the spectra. These spectra confirm that the nanocylinders with height of approximately 300nm are efficient Mie scatterers, whose modes rely on significant retardation effects along the propagation direction. Eventually, if we further increase the metamaterials height to H=450nm (Fig. 2(c)), we observe a slight increase with respect to the previous height and a shift in the extinction spectral features, in agreement with previous numerical studies.[16]

To shed light on the origin of the observed spectral features, we focus on the metamaterials with H=300nm and measure the 0[th] order transmission by narrowing the collection angular aperture to ±25° to reduce the contribution of diffuse scattering. The spectra, shown in Fig. 2(d), feature neat extinction maxima, whose spectral position spans a wide range from λ=400nm to 700nm, depending on the nanocylinder diameter. We can easily attribute them to the occurrence of dipolar Mie resonances in the nanocylinders. In fact, according to theoretical predictions, electric and magnetic dipole resonance appear close in the spectrum, approximately at the wavelength:

$$\lambda_{Mie,dipolar} = n_{eff}\, D \,,\tag{1}$$

as indicated by the vertical arrows in Fig. 2(d). In resonators with similar refractive index, the magnetic resonance has been predicted to have slightly larger wavelength than the electric one.[16] Moreover, the electric mode is expected to red-shift more quickly than the magnetic one by increasing the nanocylinder size.[15] In our measurements, we observe a pronounced peak asymmetry for all the investigated metamaterials, and even a double peak for the case D=264nm. In addition, we observe extinction maxima occurring at shorter wavelengths than the Mie resonances. Their spectral position approximately coincides with the array spacing $S$, revealing the fact that they are related to the grating condition:

$$\lambda_{grating} = S \,.\tag{2}$$

We conclude that our 2D metamaterial design supports extinction peaks due to the occurrence of Mie resonances of the nanocylinders and due to the grating condition. Their spectral position is separate, and can be independently tuned by the nanocylinder shape (mainly the diameter) and the array spacing.

To better interpret our measurements, we perform rigorous full-wave electromagnetic calculations. The scattering spectra of individual (*i.e.* isolated) nanocylinders with H=300nm are shown in Fig. 3(a). These spectra strongly resemble the measurements in Fig. 2(b), confirming that the main contribution to extinction comes from diffuse scattering. Both the calculated and measured spectra have broad spectral features, in contrast to the sharper peaks measured in Fig. 2(d). On the other hand, the calculated extinction spectra (1-T) for 2D arrays of nanocylinders are shown in Fig. 3(b). Here, we can clearly distinguish sharp extinction peaks due to dipolar Mie resonances of the nanocylinders. The identification of the resonances can be made by comparing the spectral position with the formulas in Equations 1 and 2, and by calculating the spatial distribution of the electromagnetic field. As an example, in Fig. 3(c) we show the electric and magnetic fields for the metamaterial with D=264nm excited at $\lambda_{exc}$=440nm (top) and 485nm (bottom). The first case correspond to the Mie electric dipole resonance, while the second case to the magnetic one.[15,16] The calculated peaks differ from the experimental ones in terms of their higher amplitude and of the wider spectral separation between the electric and magnetic resonances. We attribute this to three main factors. First, the effective refractive index $n_{eff}$ used in the calculations is just an approximation of the actual index, since it is derived from a homogenization method. Second, the scatterers in the fabricated metamaterials show significant geometrical imperfections, being not ideal cylinders. The top and bottom bases have different diameter and the sides are not perfectly vertical. Eventually, the substrate is neglected in the calculations (*i.e.* the 2D array is free-standing). The effects of the substrate have been already investigated, and they are well-known to determine a red-shift of the resonances and a change in the resonance amplitude.[16,29] The calculations in Fig. 3(b) also predict the onset of peaks due to the grating condition at shorter wavelengths than the Mie ones. Their spectral position is in good agreement with the experimental spectra of Fig. 2(d), due to the excellent definition in terms of array spacing in the fabrication process (see Supp. Info.). We calculate the contribution of the absorption to the extinction peak and show it in Fig. 3(d). Both the nanocylinder Mie resonances and the grating condition induce absorption peaks, and a maximum absorption value of ~50% is predicted. This is in agreement with a recent investigation on superabsorbing free-standing metamaterials, showing that 50% absorption is achieved if the forward-scattered wave is exactly $\pi$-delayed with respect of the incident one.[29] The comparison between experiments and calculations conclusively confirms the existence of absorption peaks in the investigated 2D Si-NC:SiO$_2$ metamaterials due to both the nanocylinder Mie resonances and the grating condition.

The absorption enhancement in the Si-NCs is expected to produce an increase of their photoluminescence intensity. As a matter of fact, at steady state excitation the PL intensity $I_{PL}$, expressed as photon flux, is:

$$I_{PL} \propto QY \, N_T \, \sigma \, I_{exc} \tag{3}$$

where $N_T$ is the number of Si-NCs and $I_{exc}$ is the excitation flux. The absorption cross-section is $\sigma \sim 10^{-15}$ cm$^2$ under blue excitation, a value comparable with direct-bandgap semiconductor quantum dots.[38,39] We measure the PL spectra, shown in Fig. 4(a), by exciting the samples with the 488nm line of an Ar laser. To quantify the effect of the absorption enhancement on the emission, we measure the PL intensity of the metamaterials (for the three heights H) and compare it with that of the planar Si-NC:SiO$_2$ film (of thickness equal to H) used as reference, under the same excitation conditions. We perform these measurements as function of the nanocylinder size and of the excitation wavelength. The average Si-NC concentration and QY are the same for all the samples with the same height H (including the reference), because they are fabricated starting from the same Si:NC:SiO$_2$ layer (see Methods section). However, the total number $N_T$ of emitting Si-NCs in the metamaterials is lower than the reference $\left( N_{T,ref}/N_{T,meta} = AC^{-1} \right)$. Therefore, we define the PL enhancement as:

$$PL_{enh} = \frac{I_{PL,meta}}{I_{PL,ref}} \frac{N_{T,ref}}{N_{T,meta}} = \frac{I_{PL,meta}}{I_{PL,ref}} AC^{-1} \; . \tag{4}$$

The PL enhancement is shown in Fig. 4(b) as function of the pillar parameter, for pillar height H=100 nm (celeste), 300 nm (olive) and 450 nm (honey). For the metamaterial with H=100 nm, the PL enhancement increases with the diameter and then it stays constant at a value of ~1. This simply means that the average PL intensity per Si-NC is equal to that of the reference planar film. Remarkably, for the sample with H=300 nm, $I_{PL,meta} \approx I_{PL,ref}$ despite the lower amount of Si-NCs, and the maximum PL enhancement exceeds 2 ($PL_{enh} \approx AC^{-1}$). Eventually, for the sample with H=450 nm, the PL enhancement slightly decreases from the previous case, confirming again the existence of an optimum nanocylinder height.

In order to confirm the dependence of the absorption enhancement (and therefore of the PL intensity) on the excitation wavelength, we performed photoluminescence excitation (PLE) experiments by scanning the excitation range $\lambda_{exc}$=415 – 530 nm. The results are shown in Fig. 3(c), where the circles are relative to the metamaterials with H=300nm, triangles are for H=450 nm. The metamaterial with H=300nm and D=215nm (blue line) shows an increase of PL enhancement towards short wavelengths, in agreement with the measured Mie absorption peak at $\lambda \sim$ 420nm. Analogously, the sample with D=264nm (green line), which features an absorption peak at $\lambda \sim$ 500nm, shows increased PL toward longer wavelength up to ~2.6. The sample with D=372nm (orange line) deserves a separate mention, because it features a grating condition in the investigated excitation range and, interestingly, it shows the highest enhancement value: the $I_{PL\ meta}$ / $I_{PL\ ref}$ ratio is 130% (+30%), and the nominal $AC_{meta}$ is 43%, resulting in an enhancement $PL_{enh} = 3.02$. This excellent performance is easily explained, since the fabrication method determines a small deviation on the nanocylinders separations (±2nm), which controls the grating condition, while a much larger error affects their diameter (±10nm) and shape (bases and surface roughness), reducing the amplitude of the Mie absorption peak. The metamaterials with H=450nm show analogous trends, but with lower absolute PL enhancement, as already observed in Fig. 4(b). In order to increase the enhancement factor, we could improve the quality of the nanopatterning, in terms of the nanocylinder diameter, shape and distance, and by accurately match the optical impedance of the substrate, as previously discussed in the text. However, while this effort is needed for real applications, it is not strictly necessary to demonstrate the validity of our concept.

It is important to point out that the nanopattern could modify the effective Fresnel reflection and transmission coefficients of the metamaterial samples at the emission wavelengths, thus affecting the PL angular pattern and extraction efficiency. Moreover, it could modify the photonic density of states (PDOS) and affect the Si-NC radiative lifetime, due to the Purcell effect.[40,41] In the following, we provide experimental evidences that these properties are not altered with respect to the reference planar film. A modification of Fresnel coefficients is simply detectable in the reflection and transmission spectra. Figures 5(a) and 5(b) show the variation of the reflectance $\delta R/R$ and of the transmittance $\delta T/T$, respectively, for the metamaterial samples with respect to the unpatterned film at normal incidence. In the emission spectrum, indicated by the sea-green shaded area, $\delta R/R$ and $\delta T/T$ have an average value less than 2%, which is a negligible change. This is due to the fact that the Mie resonances and the grating condition are far from the emission wavelengths. As a matter of fact, the metamaterials become more and more sub-wavelength as the emission wavelength increases, behaving as an optically-homogeneous thin film.

To directly exclude any change in the emission angular pattern, we perform Fourier (or k-space) microscopy of the PL intensity. This technique is based on capturing the Fourier plane image formed in the back focal plane of a high NA objective, which contains the k-space information of the radiative field. For all the samples, including the reference, the Fourier image is homogeneous, which indicates an isotropic PL emission. Here we focus on one representative metamaterial sample with H=300nm and D=264nm, and we first report in Fig. 5(c) the Fourier image of the transmitted light. It only shows the 0-th order central spot, which simply confirms that the metamaterial is sub-wavelength at the excitation. The Fourier image of PL is displayed in Fig. 5(d), and it shows isotropic emission. Therefore, we can exclude any effect of the PL angular dependence on the observed PL enhancement.

Furthermore, a change in the PDOS in the metamaterial samples would significantly affect the PL spectrum and lifetime. On the contrary, our measurements indicate that they do not change with the nanocylinder diameter, and they are equal to those of the reference sample. Specifically, the average PL lifetime is ~45±2 μs, as obtained by fitting the decay with a stretched-exponential function. In Fig. 5(e), we show the PL time-dynamics of the reference sample (black) and of the metamaterials with the highest PL enhancement (H=300nm and D=215nm, 264nm and 372nm). Therefore, we can exclude any effect of the PDOS on the observed PL enhancement. Eventually, Figure 5(f) shows the dependence of the PL intensity on the excitation power for the representative sample with H=300nm and D=264nm, which clearly indicates that all the samples were characterized in the linear regime, far from the saturation of the Si-NCs. We can safely attribute the increased PL in the metamaterial samples exclusively to the enhancement of the absorption at the excitation wavelength. This result was remarkably achieved without affecting the transmission/reflection properties in the emission spectrum. Moreover, the emission properties in terms of angular pattern, lifetime and spectral shape are approximately the same as the planar sample without nanocylinders.

**Conclusions.** In conclusion, we have successfully integrated quantum-confined Si-NCs into $SiO_2$ nanocylinders, arranged into a 2D metamaterial. This hierarchical metamaterial inherits the optical and electronic properties of its building blocks and shows boosted performances than the individual components. We have experimentally detected the occurrence of extinction peaks related to dipolar Mie resonances in the nanocylinders and to the grating condition, which can be tailored throughout the visible spectrum. Both of them induce spectrally-selective enhancement of the absorption, as predicted by our rigorous calculations. The Si-NCs experience increased excitation and, as a consequence, exhibit a more intense light emission. Remarkably, our experiments show that our best-performing metamaterial shows higher light emission than a planar film without nanopatterning, despite the reduced amount of Si-NCs. The metamaterial design principles described in this work are applicable to any QD semiconductor and to other emitters such as rare-earth ions and organic molecules. The specific implementation investigated in this work is completely based on Si, it is fully compatible with CMOS technology and it can be integrated with Si solar cells for spectral shaping purposes. We envision the application of this approach to photovoltaics, photocatalysis and photosynthesis.

**Corresponding Author**

* E-mail: a.capretti@uva.nl

**Author Contributions**

**Acknowledgement**

The authors acknowledge Prof. Femius Koenderink for facilitating the measurement of the metamaterial 0[th] order transmission and reflection spectra, as well of the Fourier microscopy.

**References**

(1)     Konstantatos, G.; Sargent, E. H. Nanostructured Materials for Photon Detection. *Nat. Nanotechnol.* **2010**, *5*, 391–400.

(2)     Shirasaki, Y.; Supran, G. J.; Bawendi, M. G.; Bulović, V. Emergence of Colloidal Quantum-Dot Light-Emitting Technologies. *Nat. Photonics* **2013**, *7*, 13–23.

(3)     Priolo, F.; Gregorkiewicz, T.; Galli, M.; Krauss, T. F. Silicon Nanostructures for Photonics and Photovoltaics. *Nat. Nanotechnol.* **2014**, *9*, 19–32.

(4)     Tanaka, K.; Plum, E.; Ou, J. Y.; Uchino, T.; Zheludev, N. I. Multifold Enhancement of Quantum Dot Luminescence in Plasmonic Metamaterials. *Phys. Rev. Lett.* **2010**, *105*, 1–4.

(5)     Jin, Y.; Gao, X. Plasmonic Fluorescent Quantum Dots. *Nat. Nanotechnol.* **2009**, *4*, 571–576.

(6)     Muskens, O. L.; Giannini, V.; Sanchez-Gil, J. A.; Gómez Rivas, J.; Sánchez-Gil, J. A.; Gómez Rivas, J. Strong Enhancement of the Radiative Decay Rate of Emitters by Single Plasmonic Nanoantennas. *Nano Lett.* **2007**, *7*, 2871–2875.

(7)     Soukoulis, C. M.; Wegener, M. Past Achievements and Future Challenges in the Development of Three-Dimensional Photonic Metamaterials. *Nat Phot.* **2011**, *5*, 523–530.

(8)     Jahani, S.; Jacob, Z. All-Dielectric Metamaterials. *Nat. Nanotechnol.* **2016**, *11*, 23–36.

(9)     Evlyukhin, A. B.; Novikov, S. M.; Zywietz, U.; Eriksen, R. L.; Reinhardt, C.; Bozhevolnyi, S. I.; Chichkov, B. N. Demonstration of Magnetic Dipole Resonances of Dielectric Nanospheres in the Visible Region. *Nano Lett.* **2012**, *12*, 3749–3755.

(10)    Kuznetsov, A. I.; Miroshnichenko, A. E.; Fu, Y. H.; Zhang, J.; Luk'yanchuk, B. Magnetic Light. *Sci. Rep.* **2012**, *2*, 492.

(11)    Zhao, Q.; Zhou, J.; Zhang, F.; Lippens, D. Mie Resonance-Based Dielectric Metamaterials. *Mater. Today* **2009**, *12*, 60–69.

(12)    Fu, Y. H.; Kuznetsov, A. I.; Miroshnichenko, A. E.; Yu, Y. F.; Luk'yanchuk, B. Directional Visible Light Scattering by Silicon Nanoparticles. *Nat. Commun.* **2013**, *4*, 1527.

(13)    Nieto-Vesperinas, M.; Gomez-Medina, R.; Saenz, J. J. Angle-Suppressed Scattering and Optical Forces on Submicrometer Dielectric Particles. *J. Opt. Soc. Am. A. Opt. Image Sci. Vis.* **2011**, *28*, 54–60.

(14)    Ginn, J. C.; Brener, I.; Peters, D. W.; Wendt, J. R.; Stevens, J. O.; Hines, P. F.; Basilio, L. I.; Warne, L. K.; Ihlefeld, J. F.; Clem, P. G.; *et al.* Realizing Optical Magnetism from Dielectric Metamaterials. *Phys. Rev. Lett.* **2012**, *108*, 1–5.

(15)    Staude, I.; Miroshnichenko, A. E.; Decker, M.; Fofang, N. T.; Liu, S.; Gonzales, E.; Dominguez, J.; Luk, T. S.; Neshev, D. N.; Brener, I.; *et al.* Tailoring Directional Scattering through Magnetic and Electric Resonances in Subwavelength Silicon Nanodisks. *ACS Nano* **2013**, *7*, 7824–7832.

(16)    van de Groep, J.; Polman, A.; Groep, J. Van De; Polman, A.; van de Groep, J.; Polman, A. Designing Dielectric Resonators on Substrates : Combining Magnetic and Electric Resonances. *Opt. Express* **2013**, *21*, 1253–1257.

(17)    Shcherbakov, M. R.; Vabishchevich, P. P.; Shorokhov, A. S.; Chong, K. E.; Choi, D. Y.; Staude, I.; Miroshnichenko, A. E.; Neshev, D. N.; Fedyanin, A. A.; Kivshar, Y. S. Ultrafast All-Optical Switching with Magnetic Resonances in Nonlinear Dielectric Nanostructures. *Nano Lett.* **2015**, *15*, 6985–6990.

(18)    Lin, D.; Fan, P.; Hasman, E.; Brongersma, M. L. Dielectric Gradient Metasurface Optical Elements. *Science (80-. ).* **2014**, *345*, 298–302.

(19)    Aieta, F.; Kats, M. A.; Genevet, P.; Capasso, F. Multiwavelength Achromatic Metasurfaces by Dispersive Phase Compensation. *Science (80-. ).* **2015**, *347*, 1342–1345.

(20)    Spinelli, P.; Verschuuren, M. a; Polman,  a. Broadband Omnidirectional Antireflection Coating Based on Subwavelength Surface Mie Resonators. *Nat. Commun.* **2012**, *3*, 692.

(21)    Spinelli, P.; Polman, A. Light Trapping in Thin Crystalline Si Solar Cells Using Surface Mie Scatterers. *IEEE J. Photovoltaics* **2014**, *4*, 554–559.

(22)    Trupke, T.; Green, M. A.; Würfel, P. Improving Solar Cell Efficiencies by down-Conversion of High-Energy Photons. *J. Appl. Phys.* **2002**, *92*, 1668–1674.

(23)    Kovalev, D.; Heckler, H.; Polisski, G.; Diener, J.; Koch, F. Optical Properties of Silicon Nanocrystals. **2001**, *17*, 35–40.

(24)    Timmerman, D.; Izeddin, I.; Stallinga, P.; Yassievich, I. N.; Gregorkiewicz, T. Space-Separated Quantum Cutting with Silicon Nanocrystals for Photovoltaic Applications. *Nat. Photonics* **2008**, *2*, 105–109.

(25)    Trinh, M. T.; Limpens, R.; de Boer, W. D. a. M.; Schins, J. M.; Siebbeles, L. D. a.; Gregorkiewicz, T. Direct Generation of Multiple Excitons in Adjacent Silicon Nanocrystals Revealed by Induced Absorption. *Nat. Photonics* **2012**, *6*, 316–321.

(26)    Strümpel, C.; McCann, M.; Beaucarne, G.; Arkhipov, V.; Slaoui, A.; Švrček, V.; del Cañizo, C.; Tobias, I. Modifying the Solar Spectrum to Enhance Silicon Solar Cell Efficiency-An Overview of Available Materials. *Sol. Energy Mater. Sol. Cells* **2007**, *91*, 238–249.

(27)    Švrček, V.; Slaoui, A.; Muller, J. C. Silicon Nanocrystals as Light Converter for Solar Cells. *Thin Solid Films* **2004**, *451–452*, 384–388.

(28)    Yuan, Z.; Pucker, G.; Marconi, A.; Sgrignuoli, F.; Anopchenko, A.; Jestin, Y.; Ferrario, L.; Bellutti, P.; Pavesi, L. Silicon Nanocrystals as a Photoluminescence down Shifter for Solar Cells. *Sol. Energy Mater. Sol. Cells* **2011**, *95*, 1224–1227.

(29)    Kim, S. J.; Park, J.; Esfandyarpour, M.; Pecora, E. F.; Kik, P. G.; Brongersma, M. Superabsorbing, Artificial Metal Films Constructed from Semiconductor Nanoantennas. *Nano Lett.* **2016**, acs.nanolett.6b01198.

(30)    Limpens, R.; Luxembourg, S. L.; Weeber, A. W.; Gregorkiewicz, T. Emission Efficiency Limit of Si Nanocrystals. *Sci. Rep.* **2016**, *6*, 19566.

(31)    Dohnalová, K.; Gregorkiewicz, T.; Kůsová, K. Silicon Quantum Dots: Surface Matters. *J. Phys. Condens. Matter* **2014**, *26*, 173201.

(32)    Fujiwara, H. *Principles of Spectroscopic Ellipsometry*; 2007.

(33)    Timmerman, D.; Valenta, J.; Dohnalová, K.; de Boer, W. D. a. M.; Gregorkiewicz, T. Step-like Enhancement of Luminescence Quantum Yield of Silicon Nanocrystals. *Nat. Nanotechnol.* **2011**, *6*, 710–713.

(34)    Yi, L. X.; Heitmann, J.; Scholz, R.; Zacharias, M. Phase Separation of Thin SiO Layers in Amorphous SiO/SiO 2 Superlattices during Annealing. *J. Phys. Condens. Matter* **2003**, *15*, S2887–S2895.

(35)    Miura, S.; Nakamura, T.; Fujii, M.; Inui, M.; Hayashi, S. Size Dependence of Photoluminescence Quantum Efficiency of Si Nanocrystals. *Phys. Rev. B* **2006**, *73*, 245333.

(36)    Doicu, A.; Wriedt, T.; Eremin, Y. A. *Light Scattering by Systems of Particles*; 2006; Vol. 124.

(37)    Liu, V.; Fan, S. S 4: A Free Electromagnetic Solver for Layered Periodic Structures. *Comput. Phys. Commun.* **2012**, *183*, 2233–2244.

(38)    Valenta, J.; Greben, M.; Remeš, Z.; Gutsch, S.; Hiller, D.; Zacharias, M. Determination of Absorption Cross-Section of Si Nanocrystals by Two Independent Methods Based on Either Absorption or Luminescence. *Appl. Phys. Lett.* **2016**, *108*, 23102.

(39)    Leatherdale, C. A.; Woo, W. K.; Mikulec, F. V.; Bawendi, M. G. On the Absorption Cross Section of CdSe Nanocrystal Quantum Dots. *J. Phys. Chem. B* **2002**, *106*, 7619–7622.

(40)    Novotny, L.; Hecht, B. *Principle of Nano-Optics*; Cambridge University Press, 2006; Vol. 1.

(41)    Wang, Y.; Sugimoto, H.; Inampudi, S.; Capretti, A.; Fujii, M.; Dal Negro, L. Broadband Enhancement of Local Density of States Using Silicon-Compatible Hyperbolic Metamaterials. *Appl. Phys. Lett.* **2015**, *106*, 1–5.

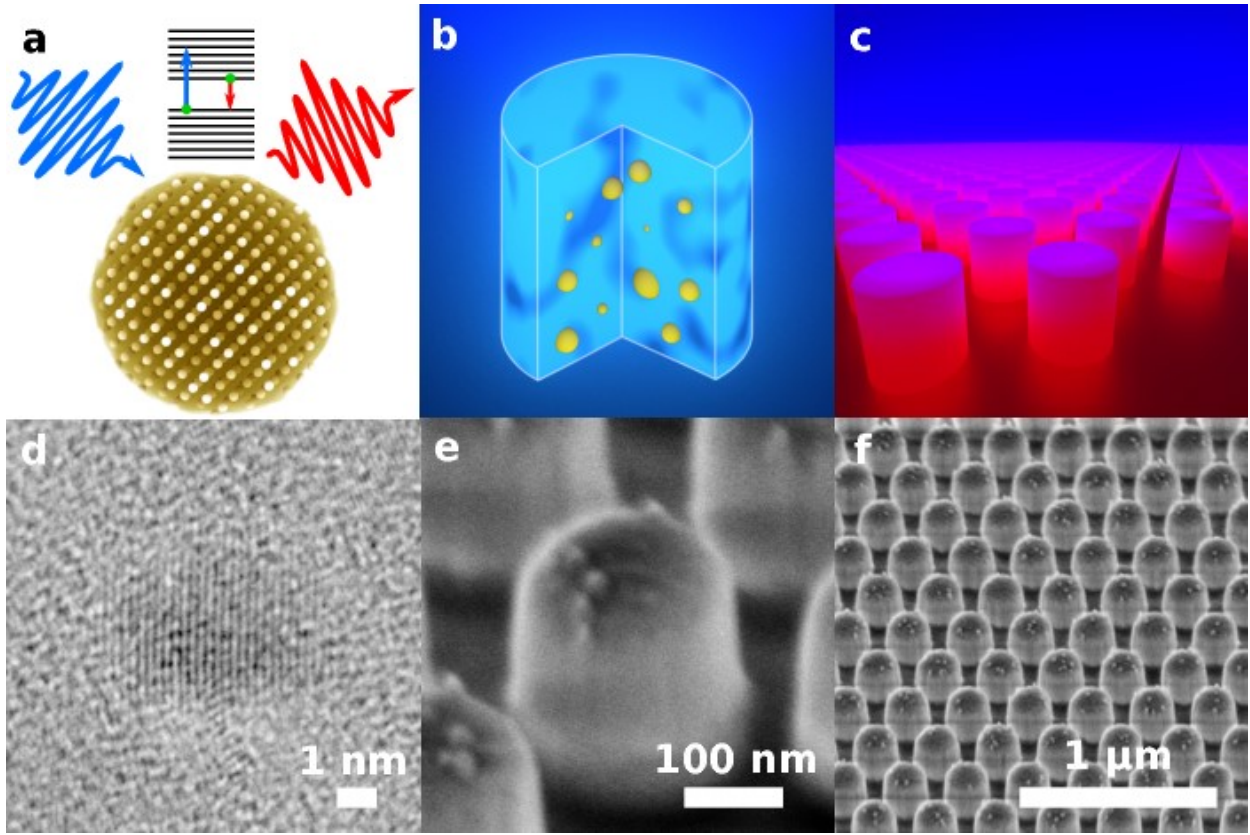

**Figure 1.** Schematic of the hierarchical metamaterial: a) quantum-confined Si-NCs, b) SiO$_2$ nanocylinders resonating at the excitation wavelength and c) photoluminescent 2D planar array. d) HRTEM of a Si-NCs in a SiO$_2$ matrix, e) SEM of a SiO$_2$ nanocylinder with height 300 nm and diameter 215 nm, and f) SEM of the 2D metamaterial. Scale bars for panels (d-f) are 1 nm, 100 nm and 1 µm, respectively.

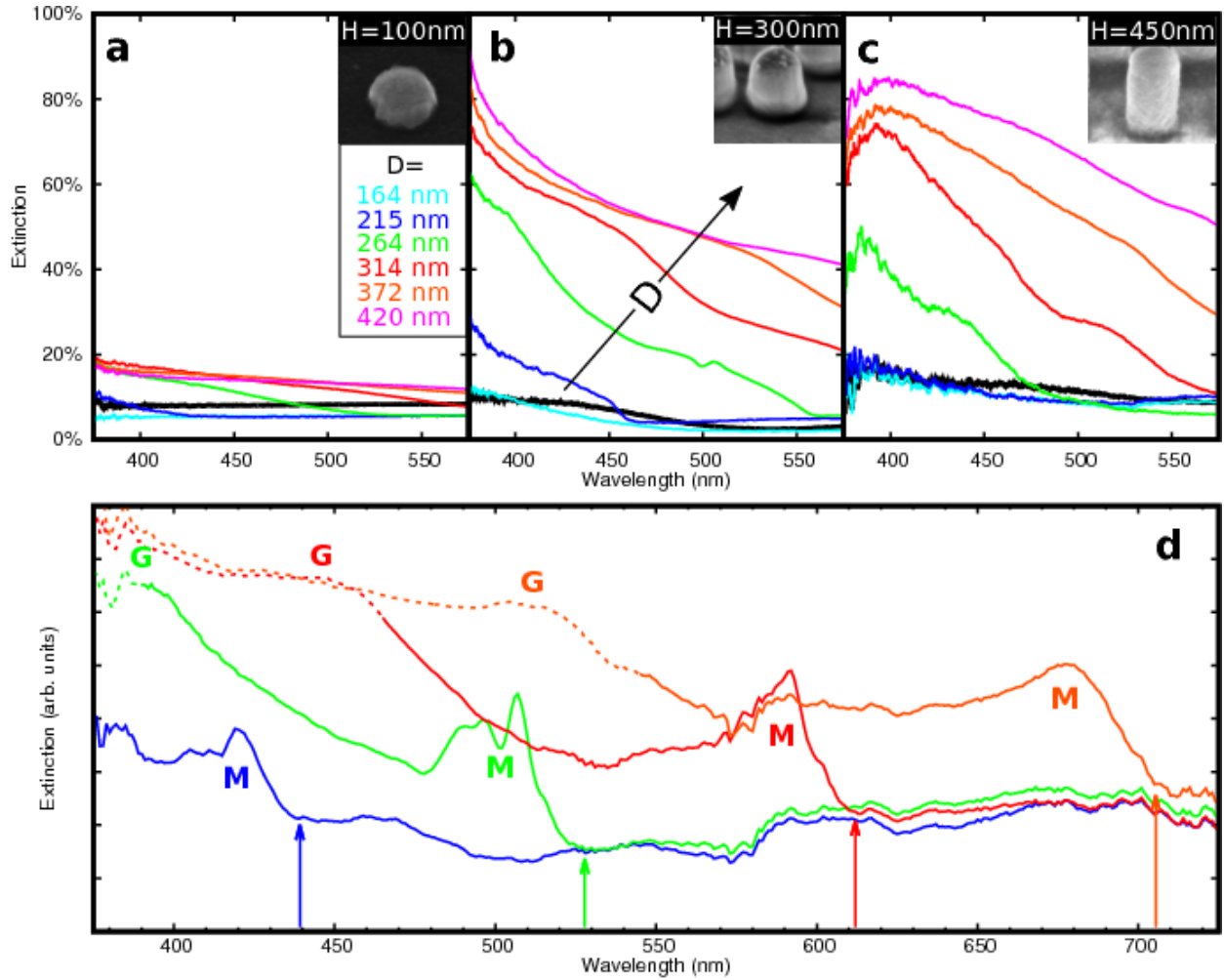

**Figure 2.** Extinction spectra (collection aperture ±49°) of metamaterials with nanocylinder height (a) H=100nm, (b) 300nm and (c) 450nm, parametrized for the diameter in the range 164 - 420 nm. The spectra of the reference planar films are reported in black for each thickness H. The insets show representative SEMs of the nanocylinder for each height. (d) 0th-order extinction spectra (collection aperture ±25°) of metamaterials with H=300nm, parametrized for the diameter from D=215 nm to 372nm. The continuous (dashed) part of the spectra indicates that the wavelength is longer (shorter) than the array spacing. Arrows indicate the predicted spectral position of the dipolar Mie resonances.

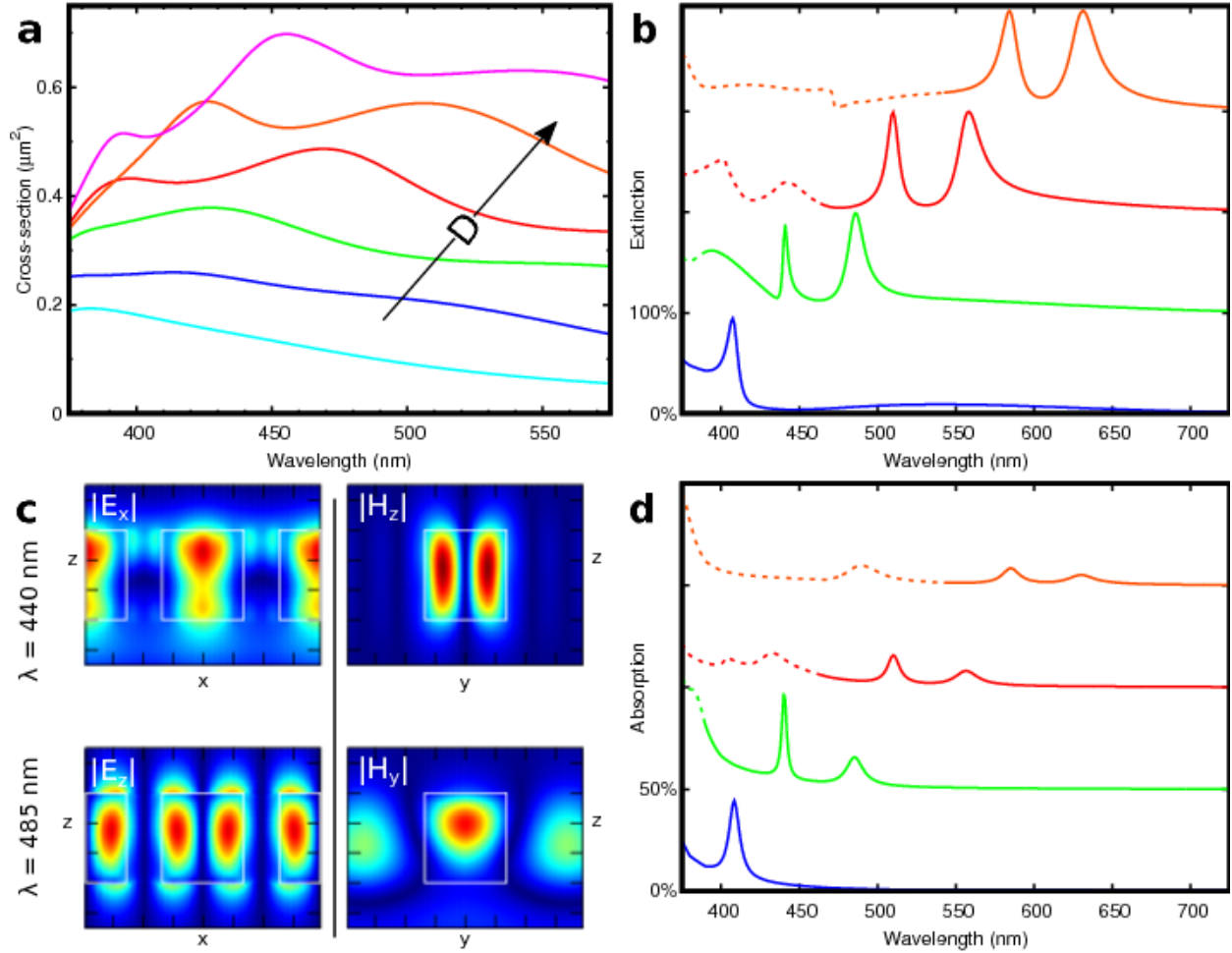

**Figure 3.** (a) Calculated scattering cross-section spectra of isolated nanocylinders parametrized for the diameter D=164nm (cyan), 215 nm (blue), 264 nm (green), 314 nm (red), 372 nm (orange) and 420 (magenta). (b) Calculated extinction spectra of a 2D array of nanocylinders parametrized for the diameter D (colors as in panel a). (c) Electric (left) and magnetic (right) field distributions for the metamaterial with D=264nm excited at $\lambda_{exc}$=440nm (top) and 485nm (bottom). (d) Same as panel (b) for absorption spectra. All the calculations are relative to nanocylinders with height H=300nm.

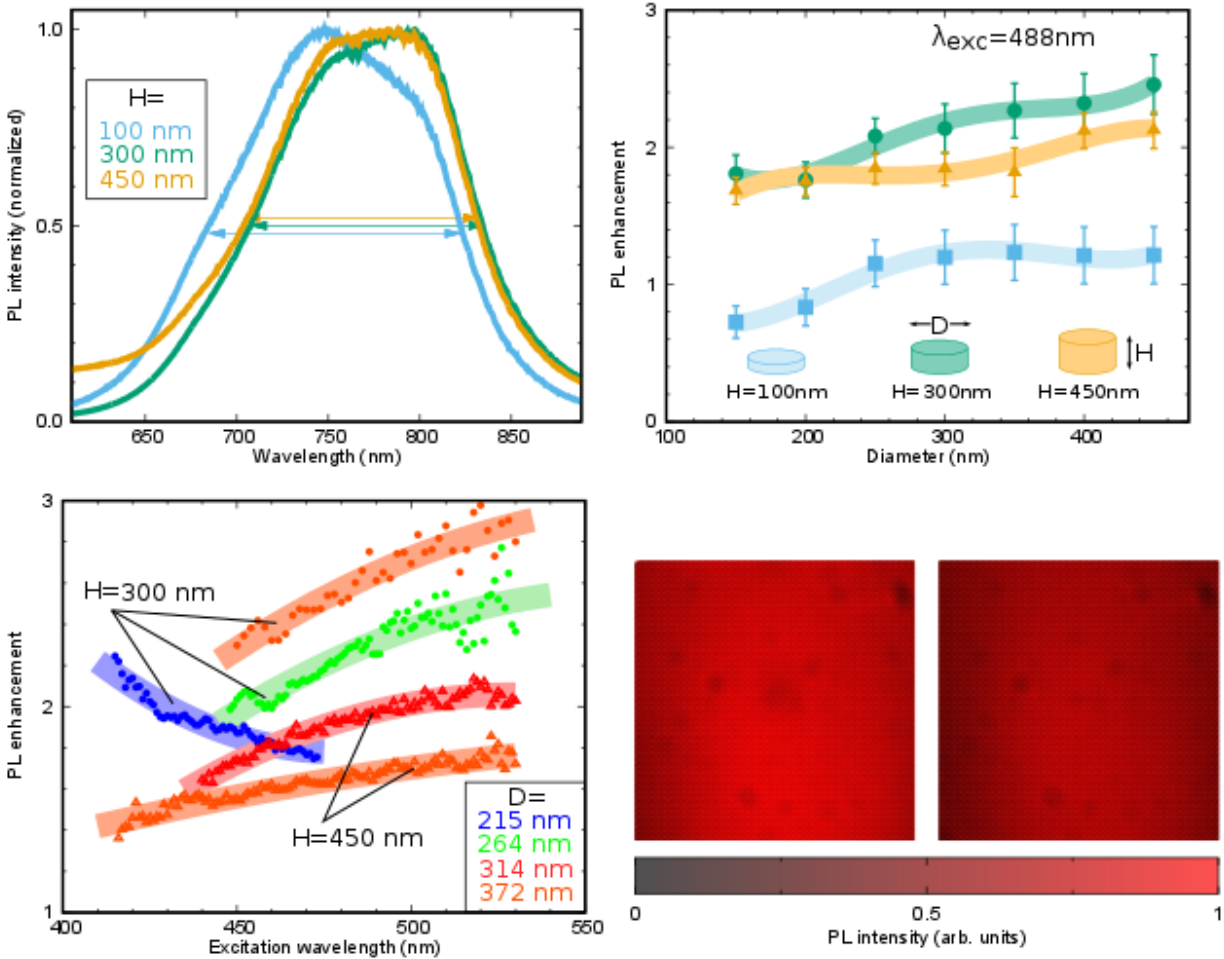

**Figure 4.** (a) Representative PL spectra of metamaterials with H=100nm (celeste), 300nm (olive) and 450nm (honey), and with D=215nm excited at $\lambda_{exc}$=488nm. (b) PL enhancement as function of nanocylinder diameter of metamaterials with H=100nm (celeste), 300nm (olive) and 450nm (honey), for $\lambda_{exc}$=488nm. (c) PLE enhancement for the metamaterial with D=215nm (blue), 264nm (green), 314nm (red) and 372nm (orange). Circles are relative to the metamaterials with H=300nm, triangles are for H=450 nm. (d) CCD image (false color) of the PL intensity from the metamaterial with H=300nm and D=372nm (left) and the reference sample (right).

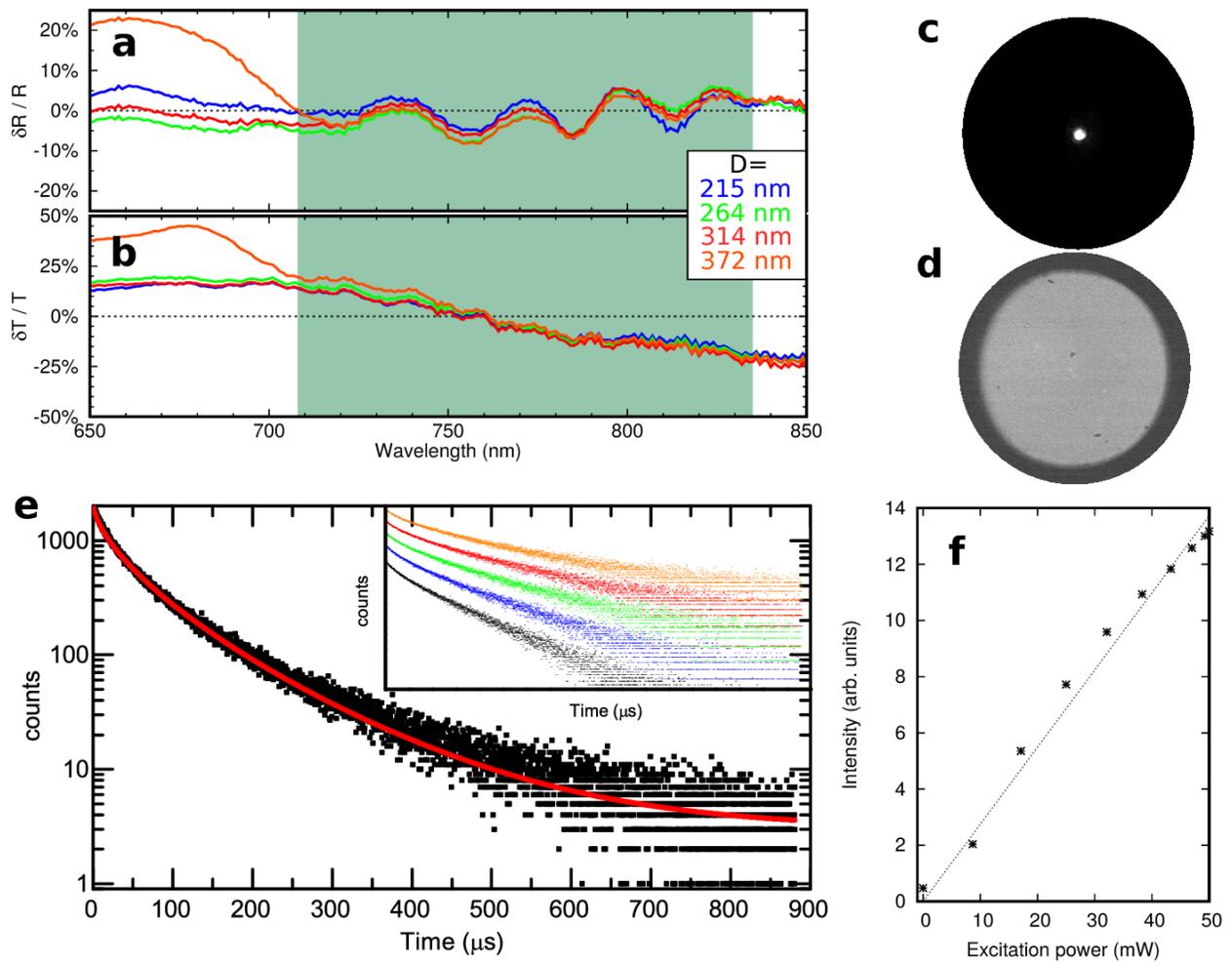

**Figure 5.** Variation of the measured reflectance (a) and transmittance (b) of the metamaterials with H=300nm and D=215nm (blue), 264nm (green), 314nm (red) and 372nm (orange), with respect to the reference. The PL FWHM spectral width is indicated by the see-green shaded area. Fourier image of the transmitted light (c) and PL (d) for the metamaterial with H=300nm and D=264nm. (e) PL intensity time-dynamics for the homogeneous reference film (black), and stretched-exponential fit (red). The inset shows the PL decays for the sample metamaterial samples as in panels (a), plotted with an artificial vertical offset, for the sake of clarity. (f) Dependence of the PL intensity on the excitation power for the same sample as in panels (c).

# Supporting Information

**Effective refractive index of the Si-NC:SiO₂ medium**

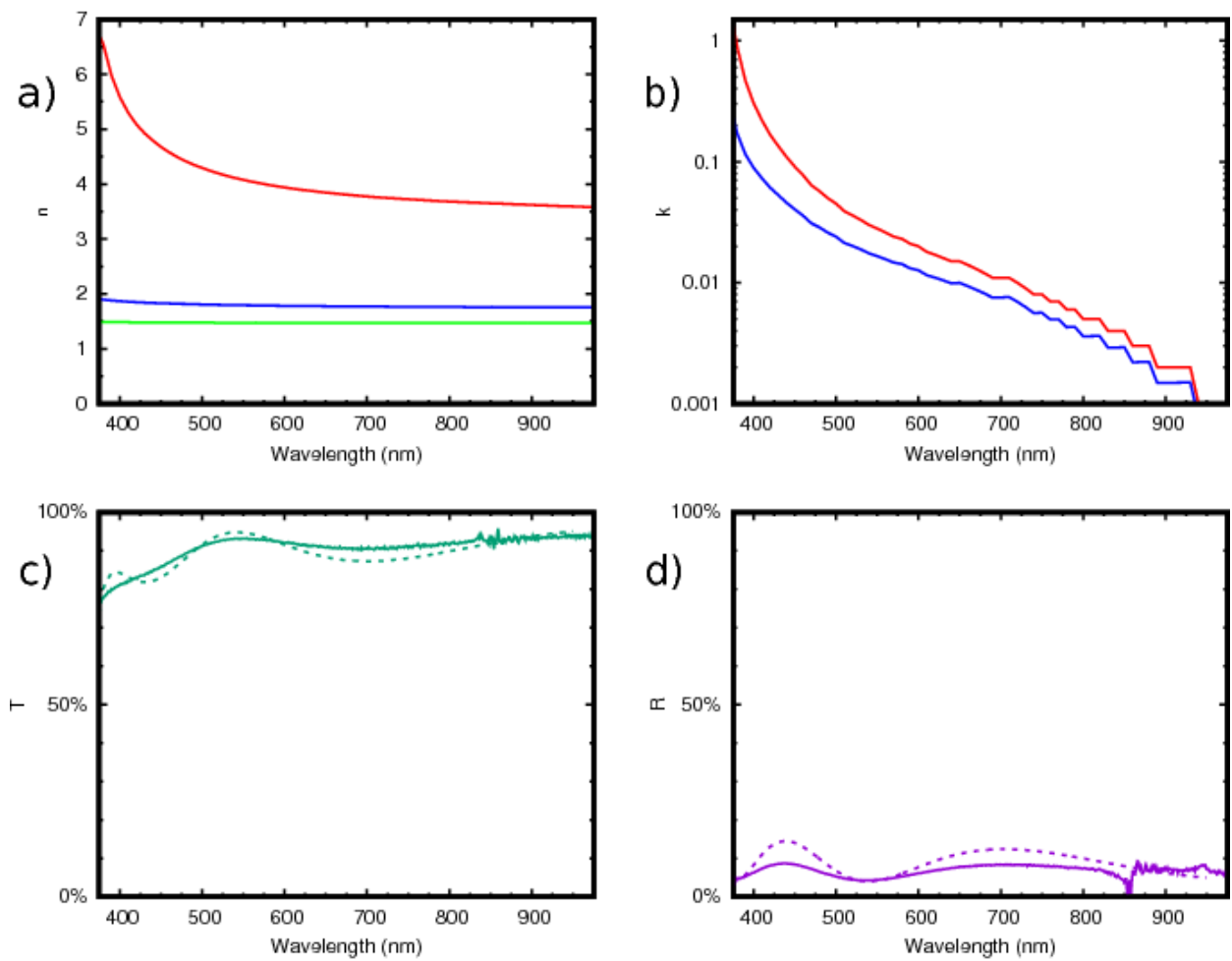

**Supporting Figure 1.** Real (a) and imaginary (b) refractive index of Si (red), SiO2 (green), and Si-NC:SiO₂ effective medium (blue). In (b), the blue curve is amplified by a factor 10 for the sake of clarity. (c) Experimental (solid line) and calculated (dashed line) transmission (a) and reflection (b) spectra of a thin film of Si-NCs in a SiO₂ matrix with thickness 300nm, on a fused silica substrate.

The real part *n* of the refractive index is shown in Supporting Figure 1(a) for Si, SiO₂ and the effective Si-NC:SiO₂ medium. The first two are taken from literature,[1,2] while the last one has been calculated using Maxwell-Garnet homogenization approach as described in the main text,[3] and has a value ~1.9 in the spectral range of interest. The imaginary part *k* of the refractive index, which indicates absorption in the medium, is shown in Supporting Figure 1(b). Optical losses in $SiO_2$ are negligible, while absorption from Si steadily increases for short wavelengths. As a result, the imaginary refractive index of Si-NC:$SiO_2$ has non-negligible absorption ($k > 0.05$) for wavelengths $\lambda < 500$nm.

Supporting Figures 1(c) and 1(d) show the transmittance and reflectance spectra, respectively, measured for a Si-NC:$SiO_2$ planar film of thickness 300nm, and the relative spectra calculated by using the well-known formula for wave propagation in stratified media.[4] The comparison reveals a good agreement between the measured optical properties and the effective index derived from the Maxwell-Garnett homogenization approach.

**Metamaterial fabrication process and list of samples**

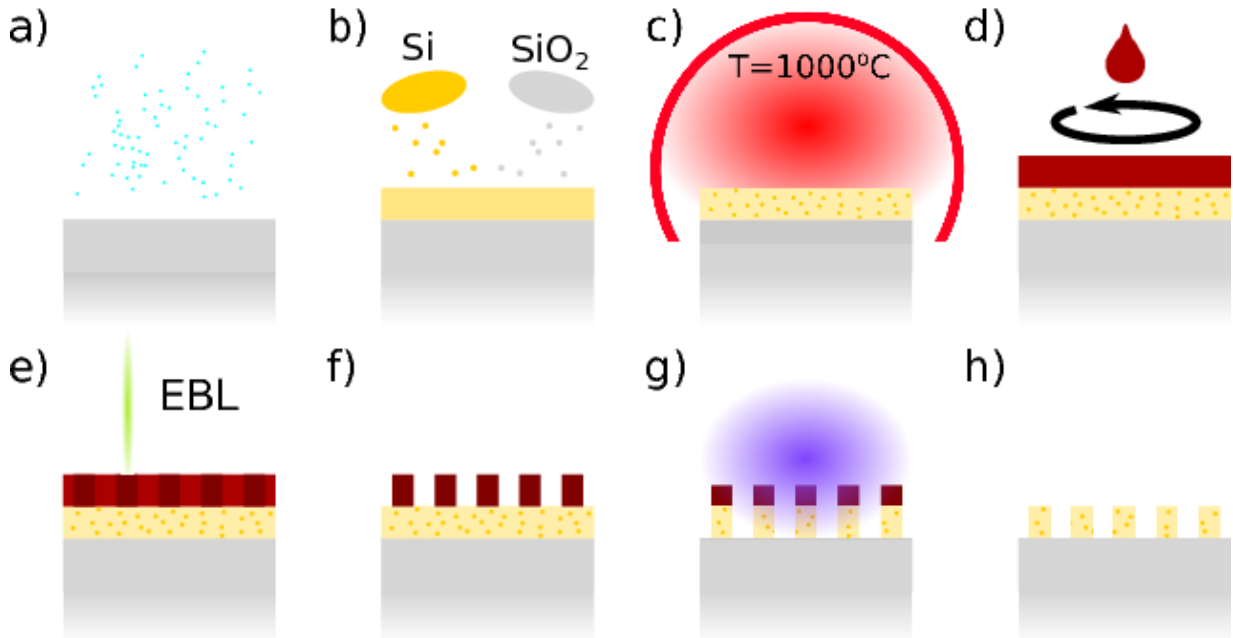

**Supporting Figure 2.** Schematic of the fabrication steps: a) substrate cleaning, b) sputtering deposition, c) thermal annealing, d) resist spin-coating, e) EBL, f) development, g) etching, h) resist removal and Si-NC passivation.

a) <u>Substrate cleaning.</u> The fused-silica substrate was sonicated in $H_2O$ and immersed in a Base Piranha solution for 15min.

b) <u>Sputtering deposition.</u> A sub-stoichiometric film of $SiO_2$ (with a 17±4 vol% excess Si) was deposited by magnetron RF co-sputtering system (AJA int.), using high purity Si (99.99%) and SiO2 (99.99%).

c) <u>Thermal annealing.</u> The sputtered film was annealed for 30 min at a temperature of 1100-1150°C in a $N_2$ environment. This resulted in both phase separation of Si and $SiO_2$, and crystallization of the Si clusters. Information on the stoichiometry and deposition is retrieved from calibrated sputter rates.

d) Resist spin-coating. A layer of HDMS primer, a thin film of MaN-2403 electron-beam resist and a layer of Espacer 300Z were consecutively spin-coated on the sample.

e) EBL. The resist is patterned by electron-beam lithography (Raith e-LiNE). The complete list of metamaterial geometries is listed in Table S1.

f) Development. The un-exposed regions of resist were removed by immersing the samples in the ma-D 525 developer.

g) Etching. The nanopattern was transferred to the underlying Si-NC:SiO$_2$ film by using an Oxford PlasmaPro100 Cobra etcher with a 10 sccm flow of C$_4$F$_8$ and of Ar.

h) Resist removal and Si-NC passivation. The remaining resist was removed by immersing the sample in acetone and by O$_2$ descum for 15min. As a final step, the sample was placed in a forming gas environment (5% H2 in N2) for 60 minutes at 500°C. This process is performed to passivate the interface between the Si-NCs and the SiO$_2$ matrix.

Slightly different sputtering and annealing paramaters were used in order to produce the three thicknesses (H=100, 300, 450nm) with the Si-NCs size distribution, which was determined to be almost identical (average diameter of ~2.5nm). The size was established from the PL spectra (see Figure 4a in the main text) as done in our past works and established literature.[5,6]

| S (nm) | D (nm) | | | | ΔAC (±%) | | |
|---|---|---|---|---|---|---|---|
| | H=100nm | H=300nm | H=450nm | Average | H=100nm | H=300nm | H=450nm |
| 230 | 158 | 168 | 168 | 164 | -5 | -1 | +4 |
| 308 | 208 | 219 | 218 | 215 | -5 | -2 | +2 |
| 388 | 260 | 272 | 261 | 264 | -6 | -3 | +1 |
| 462 | 313 | 322 | 308 | 314 | -5 | -1 | +1 |
| 543 | 368 | 374 | 374 | 372 | -6 | -1 | 0 |
| 617 | 416 | 425 | 419 | 420 | -5 | -2 | 0 |

**Supporting Table 1.** List of the metamaterial geometrical parameters for all the fabricated metamaterial height H: array spacing S, nanocylinder diameter D, and deviation ΔAC from the nominal area coverage AC=43%.

The nanopatterning process introduces some deviations between the nominal geometry and the fabricated ones. The center-to-center separation S between nanocylinders is barely affected by fabrication imperfections, and the fabricated values are equal to the nominal ones within a ±2nm error, as shown in Table S1. On the other hand, the cylinder shape is significantly affected. The upper and lower bases have different diameters and the sides are not perfectly vertical. Moreover, the nanocylinder surface shows some roughness. The fabricated diameter, as it appear from SEM analysis, is reported in Supporting Table 1. It shows significant deviation from the nominal values depending on the diameter and size. In the main text, we take into account these deviations from nominal parameter by using in Eq. 3 the fabricated values of AC.

# REFERENCES

(1)    Green, M. A.; Keevers, M. J. Optical Properties of Intrinsic Silicon at 300 K. *Prog. Photovoltaics Res. Appl.* **1995**, *3*, 189–192.

(2)    Gao, L.; Lemarchand, F.; Lequime, M. Exploitation of Multiple Incidences Spectrometric Measurements for Thin Film Reverse Engineering. *Opt. Express* **2012**, *20*, 15734–15751.

(3)    Fujiwara, H. *Principles of Spectroscopic Ellipsometry*; 2007.

(4)    Chew, W. C. *Waves and Fields in Inhomogenous Media*; IEEE Press Series on Electromagnetic Wave Theory; Wiley, 1995.

(5)    Limpens, R.; Lesage, A.; Stallinga, P.; Poddubny, A. N.; Fujii, M.; Gregorkiewicz, T. Resonant Energy Transfer in Si Nanocrystal Solids. *J. Phys. Chem. C* **2015**, *119*, 19565–19570.

(6)    Takeoka, S.; Fujii, M.; Hayashi, S. Size-Dependent Photoluminescence from Surface-Oxidized Si Nanocrystals in a Weak Confinement Regime. *Phys. Rev. B - Condens. Matter Mater. Phys.* **2000**, *62*, 16820–16825.